\documentclass[aps,prx,twocolumn,superscriptaddress,letterpaper,longbibliography]{revtex4-2}

\usepackage{graphicx}
\usepackage{bm}
\usepackage{amssymb}
\usepackage{color}
\usepackage{amsmath}
\usepackage{siunitx}
\usepackage{float}
\usepackage{enumerate}


\makeatletter
\renewcommand{\thefigure}{\@arabic\c@figure}
\makeatother
\begin{document}

\title{Versatile Braiding of Non-Hermitian Topological Edge States}

\author{Bofeng Zhu}

\affiliation{Division of Physics and Applied Physics, School of Physical and Mathematical Sciences,\\
Nanyang Technological University, Singapore 637371, Singapore}

\author{Qiang Wang}

\affiliation{Collaborative Innovation Center of Advanced Microstructures, \\School of Physics, Nanjing University, Nanjing, Jiangsu 210093, China}

\author{You Wang}

\affiliation{Division of Physics and Applied Physics, School of Physical and Mathematical Sciences,\\
Nanyang Technological University, Singapore 637371, Singapore}

\author{Qi Jie Wang}

\email{qjwang@ntu.edu.sg}

\affiliation{School of Electrical and Electronic Engineering, \\
  Nanyang Technological University, Singapore 637371, Singapore}

\author{Y.~D.~Chong}

\email{yidong@ntu.edu.sg}

\affiliation{Division of Physics and Applied Physics, School of Physical and Mathematical Sciences,\\
Nanyang Technological University, Singapore 637371, Singapore}

\affiliation{Centre for Disruptive Photonic Technologies, Nanyang Technological University, Singapore 637371, Singapore}

\email{yidong@ntu.edu.sg}

\begin{abstract}
  Among the most intriguing features of non-Hermitian (NH) systems is the ability of complex energies to form braids under parametric variation.  Several braiding behaviors, including link and knot formation, have been observed in experiments on synthetic NH systems, such as looped optical fibers.  The exact conditions for these phenomena remain unsettled, but existing demonstrations have involved long-range nonreciprocal hoppings, which are hard to implement on many experimental platforms.  Here, we present a route to realizing complex energy braids using 1D NH Aubry-Andr\'e-Harper lattices.  Under purely local gain and loss modulation, the eigenstates exhibit a variety of braiding behaviors, including unknots, Hopf links, trefoil knots, Solomon links and catenanes.  We show how these are created by the interplay between non-Hermiticity and the lattice's bulk states and topological edge states.  The transitions between different braids are marked by changes in the global Berry phase of the NH lattice.
 \end{abstract}

\maketitle

\section{Introduction}
\label{secB}

Dynamical systems lacking energy conservation, such as oscillators subject to gain or loss, can be described by non-Hermitian (NH) Hamiltonians.  Such systems need not be simple variants of their Hermitian counterparts, but can possess distinctive properties of their own \cite{Ozdemir2019, Kawabata2019, Ashida2020, Torres2020, PartoReview2021, Qiang2023}.  Research into NH Hamiltonians has unearthed successive tranches of interesting phenomena, including exceptional points (EPs) in band spectra \cite{Ozdemir2019, wiersig2020review, PartoReview2021}, parity/time-reversal and other non-Hermitian symmetries \cite{Ozdemir2019, Qiang2023}, non-Hermitian formulations of band topology \cite{Shen2018, Bergholtz2021}, the non-Hermitian skin effect \cite{Lee2016, Alvarez2018, Zhang2022}, and continua of bound states \cite{QiangCLM2023}.  Most of these phenomena stem ultimately from the absence of the spectral theorem.  For example, the NH degeneracies known as EPs involve the coalescence of eigenstates, not only eigenvalues \cite{Ozdemir2019, wiersig2020review, Bergholtz2021, PartoReview2021}.  As a related feature, parametrically encircling an EP interchanges two or more eigenstates \cite{ghosh2016ep, doppler2016dynamically}, corresponding to a closed \textit{braid} of the complex eigenvalues  \cite{Ding2022, Yang2023_2, Li2019, Luitz2019, Wu2020, Wojcik2022, Hu2022, Ryu2022, Guo2023, Konig2023, Midya2023, ZLi2023, Cao2024}.

Eigenenergy braiding cannot occur in Hermitian systems, whose energies are constrained to lie on the real line.  Through cyclic parametric variations on NH Hamiltonians, one may access topologically-distinct braids including \textit{links} and \textit{knots}, which cannot be continuously deformed into simpler braids \cite{Adams2004, Bergholtz2021, Yang2023_1, Ren2013, Li2019, Wojcik2020, Hu2021, Li2021, Wojcik2022, YLi2022, Konig2023, Zhong2023, Midya2023, Chen2024, Long2024, Long2024_2,Islam2024}.  Recently, attention has been drawn to NH bulk Hamiltonians that describe Bloch waves in periodic one-dimensional (1D) lattices.  These are indexed by the Bloch wavenumber $k$, a cyclic parameter, so the associated braiding can serve as a topological classification of NH bulk bands \cite{Wojcik2020, Hu2021, Li2021, Wojcik2022, Midya2023}.  Complex energy knots have been experimentally realized using synthetic $k$-dependent NH Hamiltonians implemented via optical fiber resonators \cite{Wang2021,Pang2024}, acoustic structures \cite{Zhang2023_1, Zhang2023_2, Qiu2023}, nitrogen-vacancy centers \cite{Yu2022}, and trapped ions \cite{Cao2023}.  There have also been recent notable experimental realizations of braiding through the careful design and encircling of EPs, such as three-band braiding by encircling order-3 EPs in optomechanical cavities \cite{Patil2022, Guria2024}, acoustic cavities \cite{Tang2022, Li2023}, and ultracold atoms \cite{Wang2023}.

In this work, we show that there is a class of NH lattices possessing topological edge states \cite{prange1990}, which exhibit a wide variety of closed braids including unknots, Hopf links, trefoil knots, Solomon links, and $[n]$-catenanes ($n\geqslant 3$) \cite{Adams2004}.  Of these, the latter two have not previously been reported in NH systems, to our knowledge.  The model in question is a NH variant of the Aubry-Andr\'e-Harper (AAH) model \cite{Harper1955, Hofstadter1976, Aubry1980, Thouless1982}, a dimensionally-reduced form of a two-dimensional (2D) quantum Hall lattice \cite{prange1990}.  We consider finite lattices with open boundary conditions, hosting not only bulk states but also topological edge states \cite{prange1990}.  Introducing non-Hermiticity induces braidings among the edge states; then, with increasing non-Hermiticity, the bulk and edge states hybridize to form a succession of more complex braids.  By comparison, previous works on energy braiding \cite{Wojcik2020, Hu2021, Wojcik2022, Wang2021,Pang2024, Zhang2023_1, Zhang2023_2, Qiu2023, Yu2022, Cao2023} have focused on \textit{bulk} NH Hamiltonians, represented by a unit cell with Floquet-Bloch boundary conditions parameterized by $k$.  Unlike the lattices we consider, these lack edges and edge states.

In Hermitian AAH lattices, a cyclic variation in $k$ can interchange topological edge states, a phenomenon tied to ``topological pumping'' in quantum Hall systems \cite{prange1990, Thouless1983, Kraus2012, Hu2017}.  For instance, Kraus \textit{et al.}~have performed experiments on laser-written AAH waveguide arrays with adiabatically varying $k$, explicitly demonstrating the evolution of an edge state into a bulk state, then back into an edge state on the opposite end~\cite{Kraus2012}.  One proviso is that in a finite-sized system, the eigenenergies for the edge states on opposite edges exhibit avoided crossings (level repulsion) at certain $k$-points where they would otherwise intersect.  As the lattice size becomes large, the ``mini gaps'' are suppressed by the exponential localization of the wavefunctions to opposite edges.

Upon introducing non-Hermiticity, the spectrum becomes complex.  The contours (complex trajectories) of the edge states' energies initially form trivial unlinked loops, or ``unlinks'' \cite{Hu2021,Wang2021}, which are continuable to Hermitian dispersion curves subject to level repulsion.  Increasing the non-Hermiticity causes the contours to link up into ``unknots'', the simplest nontrivial braid \cite{Hu2017, Hu2021, Wang2021}.  As the non-Hermiticity is further increased, the contours interact with those of the bulk states to create progressively more complex cyclic braids.

The non-Hermitian AAH model thus provides a promising setting for generating a rich variety of energy links and knots.  Compared to previously-studied NH Hamiltonians, this model presents several advantages.  First, different links and knots can be formed with merely local gain and loss; we do not require the long-range nonreciprocal or imaginary hoppings of earlier models \cite{Hu2021,Wang2021, Midya2023}, which tend to be challenging to implement experimentally.  Second, a variety of topologically distinct braids can be accessed by tuning one or two parameters; in other models, accessing different links and knots involves ad-hoc changes to the number of sites \cite{Hu2021,Wang2021,Midya2023} or the parametric trajectory \cite{Patil2022,Guria2024,Tang2022,Li2023,Wang2023}.  Third, the NH AAH model can host nontrivial braiding configurations involving significantly more eigenstates than in previous studies, which have focused on the two-band \cite{Wang2021,Zhang2023_1,Yu2022,Cao2023} and three-band case \cite{Wang2021,Li2023,Zhang2023_2,Patil2022,Guria2024,Tang2022,Wang2023}. For instance, we show that a finite lattice of length $M$ can host links belonging to a variety of $[M]$-catenane groups, including linear, circular and cyclic catenanes.

\section{Non-Hermitian Lattice Model}
\label{secC}

\begin{figure*}
  \centering
  \includegraphics[width=\linewidth]{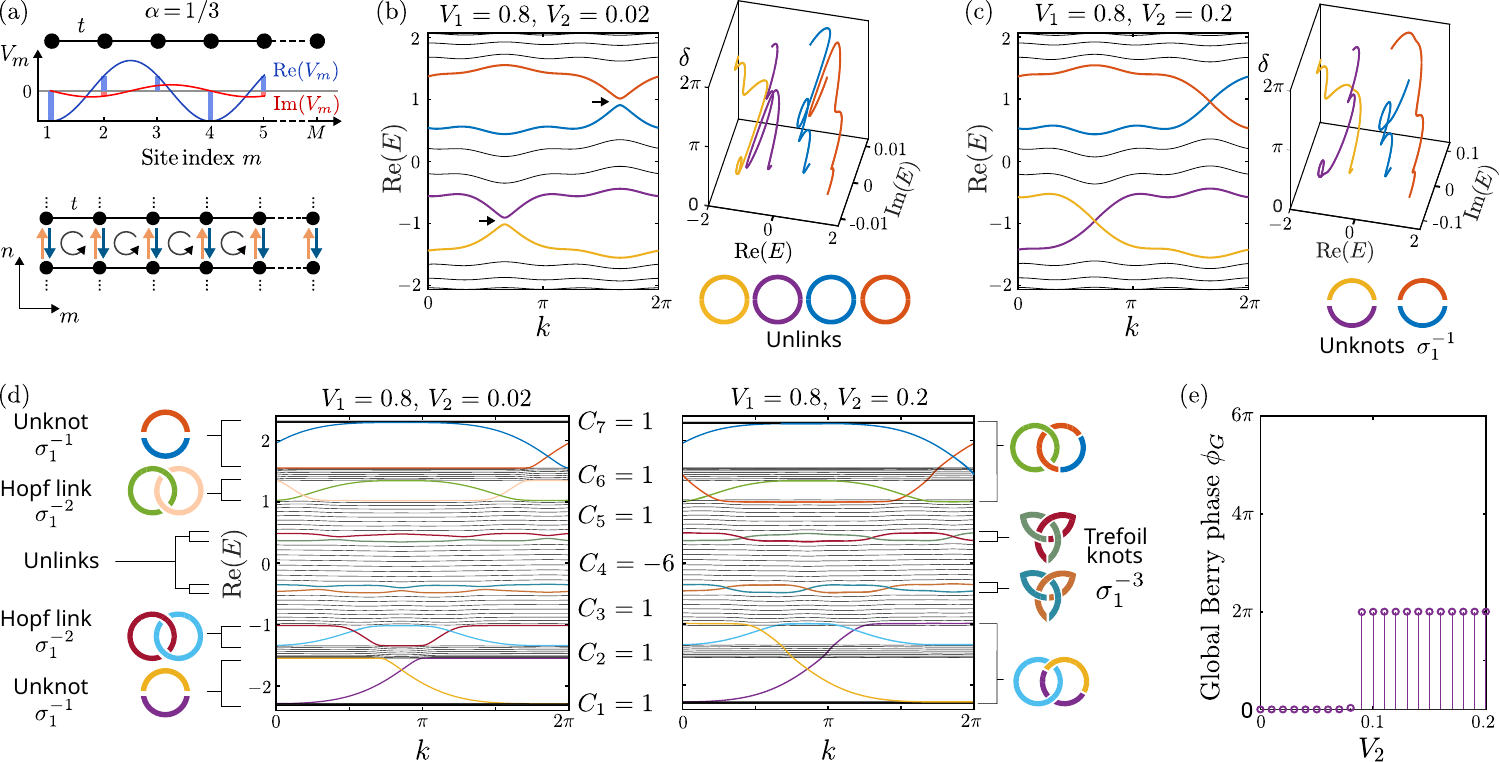}
  \caption{(a) Schematic of the NH lattice model.  Upper panel: 1D chain of length $M$ with complex on-site potential $V_m$. The real and imaginary parts of $V_m$, plotted in blue and red respectively, share an inverse period $\alpha$ and common phase $k$, with a relative phase shift of $\pi/2$. Lower panel: equivalent 2D model with asymmetric vertical hoppings (blue and orange arrows) and out-of-plane magnetic flux (curved arrows).  In this schematic, we take $\alpha = 1/3$ and $k=\pi/3$.  (b)--(c) Plot of $\mathrm{Re}(E)$ versus $k$ (left panel) and 3D plots of $\mathrm{Re}(E)$, $\mathrm{Im}(E)$ and $k$ (right panel), for two different imaginary amplitudes: (b) $V_2 = 0.02$ and (c) $V_2=0.2$.  In each case, we take $\alpha = 1/3$ and $M = 12$.  In (b), the edge state dispersion curves exhibit mini-gaps (black arrows), and the complex energy contours form unlinks, as depicted schematically below the 3D plot.  In (c), the contours have merged into unknots. (d) Plots of $\mathrm{Re}(E)$ versus $k$ for the $\alpha=1/7$ case.  The edge states form different braids for $V_2 = 0.02$ (left panel) and $V_2 = 0.2$ (right panel), including Hopf links and trefoil knots.  All other parameters are the same, including the lattice size $M=70$.  The two lattices have the same bulk NH Chern numbers $\{C_j\}$, as indicated in the space between the plots.  (e) Global Berry phase $\phi_G$ versus $V_2$ for the $\alpha = 1/3$, $M=12$ case. In all plots, we take $t = 1$.  In (b)--(e), we take $V_1=0.8$.}
  \label{fig:figure1}
\end{figure*}

We consider the 1D lattice depicted in the upper panel of Fig.~\ref{fig:figure1}(a), described by the tight-binding equations
\begin{align}
  t(\psi_{m+1}+\psi_{m-1}) + V_m\psi_m = E\psi_m,
  \label{COM_AAH_Hamiltonian_1} \\
  V_m = V_1\cos(2\pi\alpha m+k)+iV_2\sin(2\pi\alpha m+k).
  \label{Vm}
\end{align}
Here $t=1$ is the reciprocal nearest-neighbor hopping, $E \in \mathbb{C}$ is the eigenenergy, $\psi_m$ is the wavefunction, and $m = 1,\dots, M$ is the site index. The lattice is truncated to $M$ sites, with open boundary conditions.  The on-site modulation $V_m$ has both real and imaginary parts; in experimental realizations, these typically refer to the frequency detuning and gain/loss rate, respectively, of each resonator \cite{Ashida2020, Qiang2023}.  The real and imaginary parts of $V_m$ vary as cosine and sine functions of $m$, respectively, with independent amplitudes $V_1, V_2$, and the same inverse period $\alpha$ and phase $k$.  We focus on rational $\alpha=q/p$, where $p$ and $q$ are coprime positive integers, so that the bulk lattice has period $p$ \cite{Thouless1982}.

In the Hermitian limit, $V_2 = 0$, this is the famous Aubry-Andr\'e-Harper (AAH) model \cite{Harper1955, Hofstadter1976, Aubry1980, Thouless1982, Kraus2012}.  If one interprets $k$ as a Bloch wavenumber along an additional synthetic dimension, the AAH model maps to the Hofstadter model for a 2D quantum Hall (QH) lattice with out-of-plane magnetic flux density proportional to $\alpha$ \cite{Hofstadter1976, Thouless1982, prange1990, Kraus2012_2}.  The spectrum consists of bands of bulk states separated by band gaps \cite{Hofstadter1976}, which are in turn spanned by topological edge states tied to the nontrivial topology of the bulk bandstructure (nonzero Chern numbers) \cite{prange1990, Thouless1982}.

We now turn to the non-Hermitian (NH) case.  NH AAH models have been studied in a number of previous works \cite{Longhi2019PRL, Takata2018, Zhu2023}, but not in the context of energy braiding.  When $V_2$ is nonzero but relatively small, the band diagram of $\mathrm{Re}[E]$ versus $k$ is very similar to the Hermitian case, as shown in the left panel of Fig.~\ref{fig:figure1}(b).  In the complex plane, all the energy contours, including those belonging to the edge states, form separate loops or ``unlinks'' \cite{Hu2021, Wang2021}, as shown in the right panel
Fig.~\ref{fig:figure1}(b).  In the band diagram, we see that the sub-bands belonging to the edge states exhibit avoided crossings, or mini-gaps, as indicated by black arrows in the left panel of Fig.~\ref{fig:figure1}(b).  These also occur in the Hermitian case, and arise from level repulsion due to the hybridization of edge states on opposite edges.

As we further increase $V_2$, the mini-gaps shrink and eventually the unlinks turn into nontrivial closed braids called ``unknots'', as shown in Fig.~\ref{fig:figure1}(c).
Even though the edge sub-bands in the $\mathrm{Re}[E]$ plot cross each other, the actual complex contours do not intersect as they have different $\mathrm{Im}[E]$, as shown in the right panel of Fig.~\ref{fig:figure1}(c).
We can identify the braids using a ``braid word'' formed from a set of generators $\{\sigma_1,\sigma_2,...,\sigma_{M-1}\}$ \cite{Artin1947, Murasugi1999, Guo2023}: the relevant sub-bands are numbered $1,\dots,M$ in order of increasing $\mathrm{Re}(E)$, and the $i$-th band crossing over (under) the $(i+1)$-th band is denoted by $\sigma_i$ ($\sigma_i^{-1}$).  Each of the two (decoupled) unknots in Fig.~\ref{fig:figure1}(c) is described by the braid word $\sigma_1^{-1}$. The bulk eigenstates all remain as unlinks.  We note that a similar unlink-to-unknot transition for the edge states of a finite NH lattice was previously experimentally observed in a microwave network \cite{Hu2017}.


The 1D model of Eqs.~\eqref{COM_AAH_Hamiltonian_1}--\eqref{Vm} can be interpreted as a dimensional reduction of a 2D NH model, similar to the mapping between the original 1D AAH model and the 2D Hofstadter model \cite{Harper1955, Hofstadter1976, Aubry1980}.  The 2D NH model, shown in the lower panel of Fig.~\ref{fig:figure1}(a), is described by
\begin{multline}
  t(\psi_{m+1,n}+\psi_{m-1,n})
  \\+ \frac{1}{2} \sum_{\pm} (V_1\pm V_2) e^{\pm i2\pi \alpha m} \psi_{m,n\pm1}
  = E\psi_{mn},
 \label{2D_Hamiltonian}
\end{multline}
where $(m,n)$ are discrete 2D coordinates.  If the lattice is infinite along $n$, we recover the 1D model \eqref{COM_AAH_Hamiltonian_1}--\eqref{Vm}, with the wavenumber along $n$ playing the role of $k$.

For $V_2 = 0$, this reduces to the Hermitian Hofstadter model \cite{Hofstadter1976}, describing a magnetic vector potential directed along $n$ with uniform out-of-plane magnetic flux of $\alpha$ quanta per unit cell.  For $V_2 \ne 0$, the hoppings along $n$ are asymmetric: forward and backward hoppings have different magnitudes \cite{Hatano1996, Qiang2023}.  The 2D model's non-Hermiticity comes from these asymmetric hoppings, unlike the on-site gain/loss of the 1D model.

If the 2D lattice is truncated along $n$ with open boundary conditions, it exhibits the non-Hermitian skin effect \cite{Hatano1996, Longhi2015, Lee2016, Qiang2023}.  The eigenstates are exponentially concentrated on one side of the lattice, qualitatively unlike the Bloch states of the infinite lattice.  This can be explained by an imaginary gauge transformation \cite{Hatano1996, Qiang2023}, which symmetrizes the hoppings along $n$ and thereby maps the truncated NH lattice to a Hermitian Hofstadter lattice.  We will focus on the infinite-$n$ case, which has a mapping to the 1D model of Eqs.~\eqref{COM_AAH_Hamiltonian_1}--\eqref{Vm}.

Aside from the aforementioned unlink-to-unknot transition, the model can exhibit more complicated braidings.  As in the Hermitian Hofstadter model, we can vary $\alpha$ to generate different gaps and edge states \cite{Hofstadter1976}.  In Fig.~\ref{fig:figure1}(d), we show the band diagrams for $\alpha = 1/7$, with two choices of $V_2$.  For $V_2 = 0.02$, the complex sub-bands associated with the edge states form unlinks,
unknots, and Hopf links (braid word $\sigma_1^{-2}$) \cite{Hu2021, Wang2021}, as shown in the left panel of Fig.~\ref{fig:figure1}(d).  If we increase the non-Hermiticity to $V_2 = 0.2$, new braids emerge: the unknots and Hopf links hybridize, while the unlinks turn into trefoil knots ($\sigma_1^{-3}$), as shown in the right panel of Fig.~\ref{fig:figure1}(d).
Further details are given in the Supplemental Materials \cite{SM}.

Another interesting case is the $\alpha=1/9$ lattice, which is shown in Fig.~\ref{fig:figure2}(a).  We observe Solomon links formed by the edge states within a single gap ($\sigma_1^{-4}$), as well as edge states in different gaps ($\sigma_1^{-1}\sigma_2^{-2}\sigma_3^{-3}$ and $\sigma_1^{-3}\sigma_2^{-2}\sigma_3^{-1}$).

\section{Characterizing the energy braids}
\label{sec:morebraids}

How does the band topology of the 2D NH model \eqref{2D_Hamiltonian} relate to the braiding of the edge states?  Previously, it has been shown that the NH band structures of such 2D models can be characterized by integer-valued NH Chern numbers \cite{Shen2018}, similar to the Hermitian case, so long as the bulk band energies are well-separated in the complex plane.  To obtain this invariant, we define a bulk lattice cell of length $1/\alpha$, apply Floquet periodicity with wavenumber $q$, and calculate the NH Berry connection along $k$,
\begin{equation}
 A_{m}^{LR}(q,k) = i \langle \psi^L_m(q,k)|\partial_k| \psi^R_m(q,k) \rangle,
 \label{eq:A3}
\end{equation} 
where $m$ is the band index and $L/R$ denote left/right eigenvectors satisfying $\langle \psi_m^{L}(q,k)|\psi_m^{R}(q,k)\rangle = 1$.  Similar to the Hermitian case, we integrate $A_{m}^{LR}$ in momentum space using discrete Wilson loops and extract the NH Chern number from the phase windings of the eigenvalues of the Wilson loop operator \cite{Luo2019, Wang2019, Blanco2020}.

For small $V_2$, the NH Chern numbers are the same as in the Hermitian case.  For $\alpha = 1/3$ [Fig.~\ref{fig:figure1}(b)--(c)], the NH Chern numbers are $C_1 = 1, C_2 = -2, C_3 = 1$, in order of increasing $\mathrm{Re}(E)$.  According to the bulk-boundary correspondence, the number of topological edge states per edge per gap is $\sum_{j}C_j$, where $j$ is summed over all bands below the gap, and this is consistent with the NH band spectra of Fig.~\ref{fig:figure1}(b)--(c).  Hence, the NH Chern number predicts the multiplicity of edge state sub-bands available for braiding in each bandgap, but does \textit{not} determine the actual braidings.  For example, the unlinks in Fig.~\ref{fig:figure1}(b) and unknots in Fig.~\ref{fig:figure1}(c) occur under the same NH Chern numbers.  Similarly, in Fig.~\ref{fig:figure1}(d) the NH Chern numbers are the same for $V_2 = 0.02$ (left panel) and $V_2 = 0.2$ (right panel), but the edge state braidings are different.

We next consider an alternative topological invariant, the global Berry phase \cite{LiangPRA2013, Takata2018, Parto2018}. For a finite chain with open boundary conditions, with $k$ as a tunable parameter, the global Berry phase is defined as \cite{LiangPRA2013}
\begin{align}
  \phi_G &= \sum_{m=1}^{M} \int_0^{2\pi} A_{m}^{LR}(k) \,dk, \label{eq:A2}
\end{align}
where $A_{m}^{LR}(k)$ is given by Eq.~\eqref{eq:A3}, with no $q$-dependence (as we are using open and not Floquet-periodic boundary conditions).  Since a gauge transformation can change $\phi_G$ by multiples of $2\pi$, we identify phase transitions by abrupt changes in $\phi_G$ under a given gauge.  To fix the gauge, we take the first component of $|\psi^{L/R}_m(k)\rangle$ to be real and positive.

\begin{figure}
  \centering
  \includegraphics[width=\linewidth]{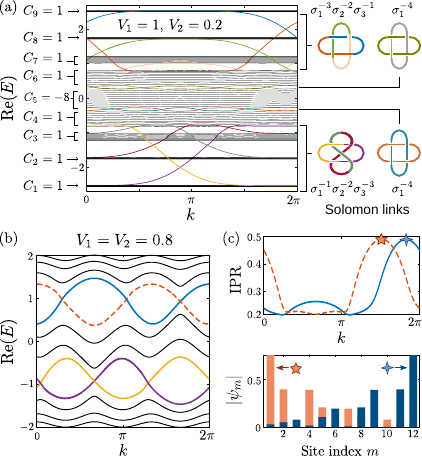}
  \caption{(a) Band diagram for the $\alpha = 1/9$ lattice, with parameters $M = 99$, $V_1 = 1$, and $V_2 = 0.2$.  The bulk bands are well-separated and the NH Chern numbers $C_1, \dots, C_9$ are indicated to the left of the plot.  The edge states form Solomon links.  (b) Band diagram for the lattice of $V_1 = V_2 = 0.8$, with all other parameters the same as in Fig.~\ref{fig:figure1}(b)--(c).  Four of the complex energy contours, plotted in blue/orange and yellow/purple, form a pair of trefoil knots.  (c) Upper panel: plot of the the inverse participation ratio (IPR), a measure of localization \cite{Thouless1974}, versus $k$ for the upper (blue/orange) set of contours in (b).  Lower panel: spatial distribution of the maximally-localized eigenfunctions indicated by the stars in the upper panel.  }
  \label{fig:figure2}
\end{figure}

The global Berry phase is known to give results consistent with the spectral winding numbers of complex energy contours \cite{Wang2021, Zhang2023_1, Zhang2023_2, Midya2023}.  In particular, Hu and Zhao \cite{Hu2021} have identified a relationship between $\phi_G$ and the band permutation index for a two-band model.  We find that a generalization of this relationship applies to the NH AAH model.  Given a braid word
\begin{equation*}
  \prod_{m=1}^{M-1} \sigma_{m}^{-n_m},
\end{equation*}
the global Berry phase is determined by the braid's ``crossing number'' \cite{Murasugi1999, Zhong2023}, as follows:
\begin{equation}
  \phi_G = |\pi \sum_{m=1}^{M-1} n_m|.
  \label{phig_braid}
\end{equation}
Moreover, decoupled braids contribute additively to $\phi_G$.  We have verified that this relation holds for all the braid words described in this work.

In Fig.~\ref{fig:figure1}(e), we plot $\phi_G$ versus the imaginary modulation amplitude $V_2$ for $\alpha = 1/3$ and $M=12$.  At $V_2 \approx 0.08$, coincident with the braiding transition between Fig.~1(b) and (c), there is a jump from $\phi_G=0$ to $\phi_G=2\pi$, in agreement with Eq.~\eqref{phig_braid} with each unknot contributing $\pi$.  Likewise, for the band diagrams in Fig.~1(d), we find $\phi_G =6\pi$ for $V_2 = 0.02$ (two sets of unknots and Hopf links), and $\phi_G=60\pi$ for $V_2 = 0.2$ (multiple braids, including contributions from bulk sub-bands).  For details, see the Supplemental Materials \cite{SM}.

Upon further increasing the non-Hermiticity, more complicated behaviors emerge.  In Fig.~\ref{fig:figure2}(b), we show the band diagram for $\alpha = 1/3$ and $V_1 = V_2 = 0.8$.  Compared to Fig.~\ref{fig:figure1}(c), the increase in $V_2$ has created additional crossings of the real line gaps.  Accordingly, the two unknots in Fig.~\ref{fig:figure1}(c) have turned into two trefoil knots in Fig.~\ref{fig:figure2}(b).  (All the other contours remain as unlinks.)

\begin{figure*}
  \centering
  \includegraphics[width=\linewidth]{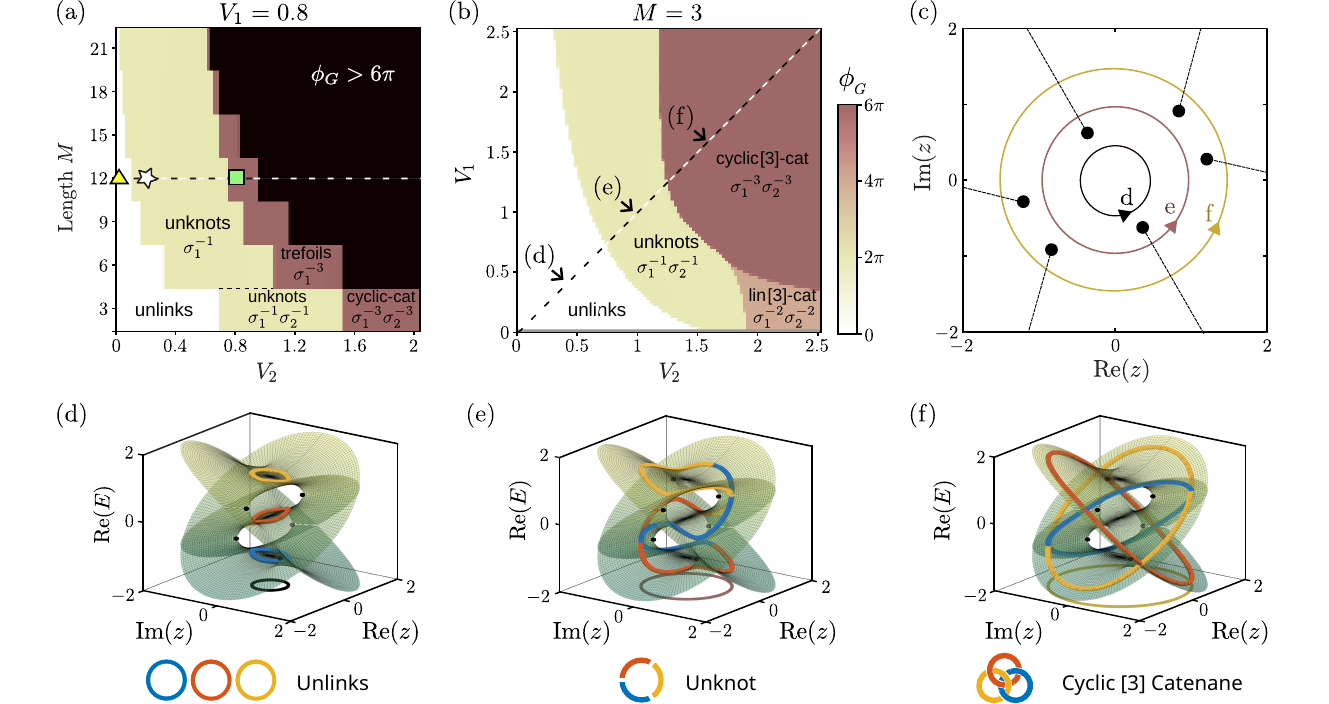}
  \caption{(a) Phase diagram for energy braids, expressed in terms of $M$ and $V_2$, for fixed $\alpha = 1/3$ and $V_1 = 0.8$.  The lattice size $M$ is incremented in steps of 3.  The colors indicate the global Berry phase $\phi_G$, and the braid words in several areas is indicated; the white area ($\phi_G > 6\pi$) contains numerous complicated braids. The triangle, star, and square markers correspond to Fig.~\ref{fig:figure1}(b), Fig.~\ref{fig:figure1}(c), and Fig.~\ref{fig:figure2}(b) respectively. For large $M$, the trefoil phase persists over a successively smaller range of $V_2$. (b) Phase diagram in terms of $V_1$ and $V_2$ for the $\alpha = 1/3$, $M = 3$ trimer.  The points labeled (d)--(f) correspond to the respective subplots below.  (c) EPs (black dots) for the trimer with $V_1 = V_2$, expressed in the parameter space $z = V_{1} e^{ik}$.  The circles labelled (d)--(f) are parametric trajectories for $k \in [0, 2\pi)$, which correspond to the following subplots and encircle zero, two, and six EPs respectively.  (d)--(f) Riemann surface trajectories for (d) $V_1 = V_2 = 0.4$, (e) $V_1 = V_2 = 1$ and (f) $V_1 = V_2 = 1.6$. The color solid curves on the Riemann surfaces denote the trajectories of lattice eigenvalues Re$(E)$ when $k$ varies from $0$ to $2\pi$. The projections of these trajectories on the complex plane are also illustrated. The solid dots are EPs. }
  \label{fig:figure3}
\end{figure*}

Although these additional gap-crossings seem reminiscent of additional edge states, they turn out \textit{not} to be localized to the edge.  The upper panel of Fig.~\ref{fig:figure2}(c) shows the inverse participation ratio (IPR) $\sum_{m} |\psi_{m}|^4 / (\sum_{m} |\psi_{m}|^2)^2$, a standard measure of localization \cite{Thouless1974}, for the upper pair of energy contours (colored blue and orange) in Fig.~\ref{fig:figure2}(b) that form a trefoil knot.  As we vary $k$, we see two IPR peaks at $k \approx 1.7\pi$, where the topological edge states originally crossed the gap.  The wavefunctions are indeed localized to opposite edges of the lattice, as shown in the lower panel of Fig.~\ref{fig:figure2}(c).  The new gap-crossings do not give additional IPR peaks, and their states are extended across the lattice. More details are given in the Supplementary Materials \cite{SM}.

\section{Phase diagrams and exceptional points}
\label{secE}


Using the global Berry phase, we can produce phase diagrams for the complex energy braids.  In Fig.~\ref{fig:figure3}(a), we plot the phase diagram in terms of $M$ and $V_2$, for fixed $\alpha = 1/3$ and $V_1 = 0.8$.  The lattice length $M$ is incremented in steps of 3, so that the truncation conditions remain the same.  The diagram contains an assortment of phases, labeled by their braid words.  These include the unlink phase of Fig.~\ref{fig:figure1}(b) (marked by a triangle), the unknot phase of Fig.~\ref{fig:figure1}(c) (marked by a star), and trefoil knot phase of Fig.~\ref{fig:figure2} (marked by a square).

In viewing the phase diagram, it is worth bearing in mind that (i) distinct braids can have the same $\phi_G$, based on Eq.~\eqref{phig_braid}, and (ii) the same link/knot name can correspond to different braid words, as previously seen in Fig.~\ref{fig:figure2}(a).  For instance, the $\phi_G = 2\pi$ region in Fig.~\ref{fig:figure3}(a) contains two unknot phases: $\sigma_1^{-1}$, consisting of two decoupled unknots formed by two pairs of contours, and $\sigma_1^{-1}\sigma_2^{-1}$, consisting of one unknot formed by three contours [see Fig.~\ref{fig:figure3}(e)].  Similarly, the $\phi_G = 6\pi$ region contains a $\sigma_1^{-3}$ trefoil knot phase and a $\sigma_1^{-3}\sigma_2^{-3}$ cyclic [3]-catenane phase.

\begin{figure*}
  \centering
  \includegraphics[width=\linewidth]{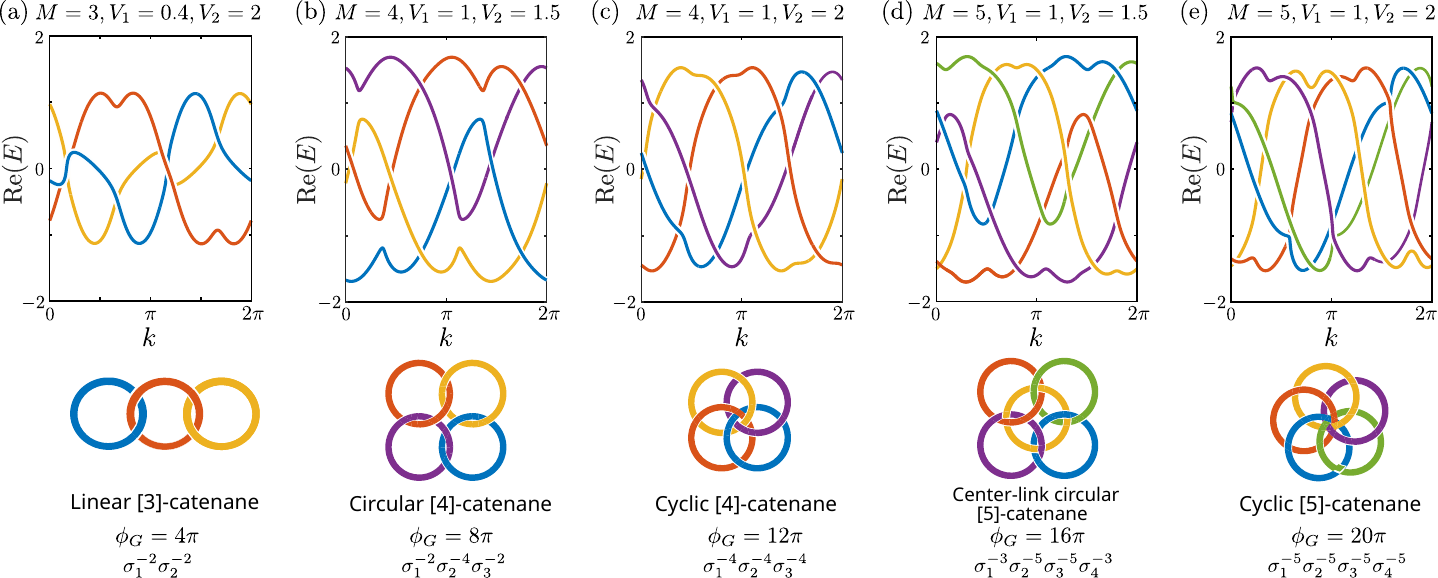}
  \caption{Examples of $[n]$-catenanes in lattices with (a) $\alpha = 1/3$, (b)--(c) $\alpha = 1/4$, and (d)--(e) $\alpha = 1/5$.  In all cases, $M = 1/\alpha$.  Upper panels show the band diagrams, and lower panels show the knot, the global Berry phase $\phi_G$, and the braid word.}
  \label{fig:figure4}
\end{figure*}

Distinct braiding phases occur even in lattices as small as $M = 3$, i.e., trimers.  We will study this case in more detail, as it serves as a prototype for how braiding transitions occur.  (Many braiding behaviors in larger lattices evolve from the trimer; see Appendix~\ref{appendix:latticesize}.)  Fig.~\ref{fig:figure3}(b) shows a phase diagram for trimers, in terms of $V_1$ and $V_2$.

Consider the special case of $V_1=V_2$, marked by the dashes in Fig.~\ref{fig:figure3}(b).  For this case, the Hamiltonian is
\begin{equation} \label{Trimer_finite}
 \mathcal{H}(z) =
 \begin{bmatrix}
   z e^{2\pi i/3} & t & 0\\
   t & z e^{-2\pi i/3} & t\\
   0 & t & z \\
 \end{bmatrix},
\end{equation}
where $z = V_{1} e^{ik}$.  Over one cycle of $k \in [0, 2\pi)$, the complex variable $z$ executes a circular parametric loop.  The eigenenergies are solutions to the cubic equation
\begin{equation*}
  -E^3 + 2t^2 E + e^{-2\pi i/3} t^2z + z^3 = 0,
\end{equation*}
which has double roots when
\begin{equation} \label{double_roor_criteria}
  \begin{aligned}
    27 z^2 (e^{-2\pi i/3} t^2 + z^2)^2 - 32 t^6 = 0.
  \end{aligned}
\end{equation}
The solutions are six EPs of the form $\{\pm z_1, \pm z_2, \pm z_3\}$, where $|z_1| = |z_2| > |z_3|$, as shown in Fig.~\ref{fig:figure3}(c).

Over one cycle of $k$, if the parametric trajectory does not intersect with any EP, it must encircle either zero, two, or six EPs, depending on the choice of $V_1$.  This accounts for the three distinct energy braiding behaviors observed in these trimers.  If zero EPs are encircled, the contours on the Riemann surfaces form separate loops and the energies form unlinks, as shown in Fig.~\ref{fig:figure3}(d).  Encircling two EPs, which is equivalent to encircling a third-order EP, produces a cyclic permutation of the three energies and hence an order-3 unknot, as shown in Fig.~\ref{fig:figure3}(e).  If six EPs are encircled, the contours on the Riemann surfaces are nested to create a cyclic $[3]$-catenane, as shown in Fig.~\ref{fig:figure3}(f).

As shown in Appendix \ref{secH}, a similar EP-encircling argument can be applied to dimers with $M = 2$, $\alpha = 1/2$, to explain the transition from unlinks to Hopf links.

Although the connection between braiding and EP encircling has previously been explored \cite{Wang2021, Zhang2023_1, Yu2022, Cao2023, Li2023, Zhang2023_2, Patil2022, Guria2024, Tang2022, Wang2023, Qiu2023, Pang2024}, these studies focused on braids of two or three complex energies.  Building on the behavior of the NH trimer, we can generate more complicated braids while maintaining a link to EP encircling.  For example, lattices of length $M$ and inverse period $\alpha=1/M$ host a variety of [$M$]-catenanes \cite{Stoddart2020}.  The $M = 3$ lattices host cyclic $[3]$-catenanes [Fig.~\ref{fig:figure3}(f)], and also linear $[3]$-catenanes [Fig.~\ref{fig:figure4}(a)]; the latter corresponds to the $\phi_G = 4\pi$ phase in the lower-right corner of Fig.~\ref{fig:figure3}(b).  For $M = 4$, we can generate circular $[4]$-catenanes and cyclic $[4]$-catenanes, as shown in Fig.~\ref{fig:figure4}(b)--(c).  For $M = 5$, we can generate center-link circular $[5]$-catenanes and cyclic $[5]$-catenanes, as shown in Fig.~\ref{fig:figure4}(d)--(e).  Even more configurations can be found by varying $V_1$ and/or $V_2$.

\section{Discussion}
\label{secF}

We have presented a family of non-Hermitian lattices based on the Aubry-Andr\'e-Harper (AAH) model, whose complex eigenenergies exhibit topologically distinct braids.  Previously-studied braiding models have required nonreciprocal long-range couplings, which are challenging to realize experimentally and can only be reliably implemented on a handful of specialized experimental platforms \cite{Hu2021, Wang2021, Midya2023}.  By contrast, our model requires only reciprocal nearest-neighbor couplings, along with on-site gain/loss modulations.  It should thus be significantly easier to implement experimentally.  For example, there have been previous experimental demonstrations of coupled-cavity lasers with varying resonance frequencies and gain/loss on individual cavities \cite{Brandstetter2014, Zhao2018}.  


The model can access a wide variety of braiding configurations, including trefoil knots, Solomon links, and various linear and cyclic $[n]$-catenanes.  Several of these links and knots have not previously been reported in lattice models, even theoretically.  These braids do not require large lattices; for example, lattice sizes of 3, 4, and 5 are sufficient to host many interesting $[n]$-catenanes (Fig.~\ref{fig:figure4}).

The cyclic parameter $k$ that drives the braidings is the phase of a complex modulation function, which maps to the wavenumber in a non-Hermitian Hofstadter model.  This thereby establishes a link between non-Hermitian eigenvalue braiding and the physics of 2D non-Hermitian topological insulators \cite{Shen2018, Bergholtz2021}. Moreover, we have found a correspondence between the braid words and the global Berry phase, a non-Hermitian topological invariant calculated from the eigenstates of the lattice Hamiltonian \cite{LiangPRA2013, Takata2018, Parto2018}.

As the braids are discrete phenomena (which is also reflected in the quantization of the global Berry phase), they are unchanged by weak disorder in the lattice.  However, sufficiently strong disorder can change the braids.  This includes the interesting possibility of generating braids that do not occur in the ordered lattice, an example of which is given in Appendix~\ref{appendix:EPdisorder}.

We have also shown how the transitions may be understood in terms of exceptional point encirclings in a complex parameter space.  In the future, it may be desirable to use this line of reasoning to analyze other braids, including those in larger lattices.  Such an understanding might be helpful for designing non-Hermitian systems with deliberately-chosen braiding configurations, which may eventually be useful for signal modulation schemes and device applications.

We are grateful to Haoran Xue for helpful discussions. This work was supported by the Singapore National Research Foundation (NRF) Competitive Research Program (CRP) Nos.~NRF-CRP23-2019-0005, NRF-CRP23-2019-0007, and NRF-CRP29-2022-0003, the NRF Investigatorship NRF-NRFI08-2022-0001 and Singapore A*STAR Grant No.~A2090b0144.

\appendix

\section{Effects of lattice size on braiding}
\label{appendix:latticesize}

\begin{figure}
  \centering
  \includegraphics[width=\linewidth]{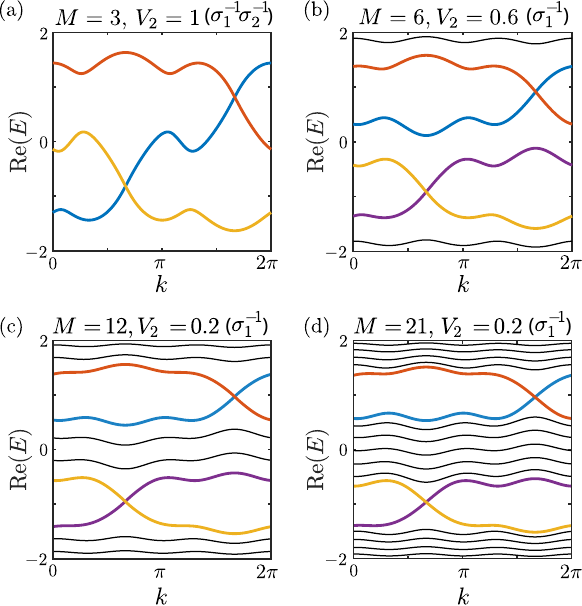}
  \caption{Band diagrams for $\alpha = 1/3$ lattices of different $M$, with different values of $V_2$ chosen to lie in the $\phi_G = 2\pi$ (unknot) phase, according to the phase diagram in Fig.~\ref{fig:figure3}(a).  In all subplots, we take $V_1 = 0.8$.}
  \label{fig:figureS3}
\end{figure}

For fixed inverse period $\alpha$, different lattice lengths $M$ (in increments of $1/\alpha$) can produce similar braiding behaviors.  We saw this in Fig.~\ref{fig:figure3}(a), where the phase diagram is plotted with $M$ on one axis, and several phases are found to extend over multiple $M$.  In particular, the $\phi_G = 0$ unlinks phase and the $\phi_G = 2\pi$ unknots phase both extend from $M = 3$ to large $M$.

As a further illustration, Fig.~\ref{fig:figureS3} shows band diagrams for different $M$, in the $\phi_G = 2\pi$ unknots phase.  Fig.~\ref{fig:figureS3}(a) features an order-3 unknot (braid word $\sigma_1^{-1}\sigma_2^{-1}$) for the $M=3$ lattice, which we have previously shown to be equivalent to encircling a pair of order-2 EPs (see Sec.~\ref{secE}). In Fig.~\ref{fig:figureS3}(b), the lattice size is increased to $M = 6$, which introduces an additional band.  As a result, the order-3 unknot breaks up into a pair of order-2 unknots  (each having braid word $\sigma_1^{-1}$).  Increasing $M$ adds even more sub-bands, and the band diagram eventually comes to resemble that of a topological insulator, with braiding between the topological edge states that span the two bulk band gaps.

\begin{figure}
  \centering
  \includegraphics[width=\linewidth]{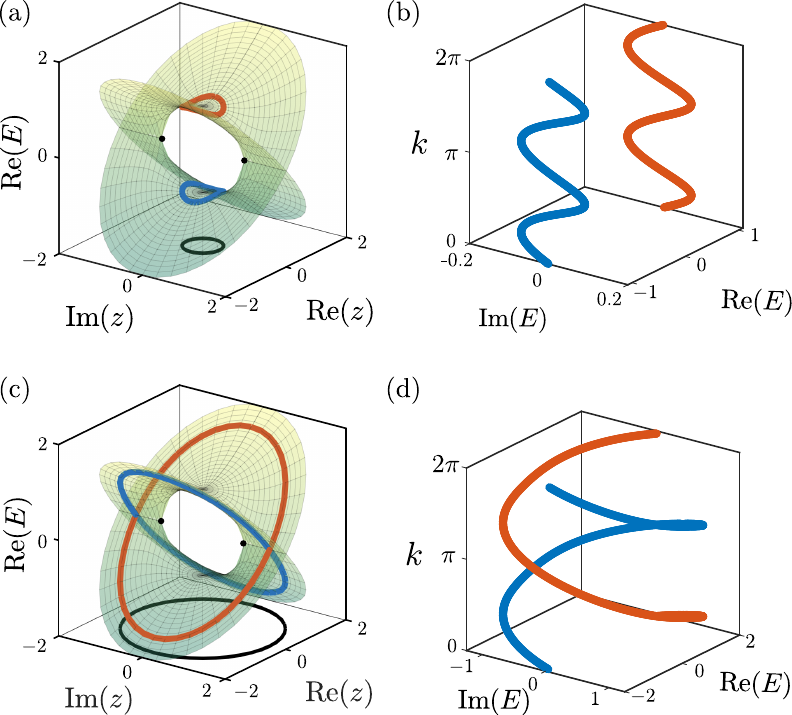}
  \caption{(a) Riemann surfaces for the $M = 2$, $\alpha = 1/2$, $t = 1$ dimer, plotted in terms of $z = V_1 \mathrm{cos}(k) + iV_2 \mathrm{sin}(k)$.  The black dots indicate the order-2 EPs, and the red and blue curves correspond to the parametric trajectory over one cycle of $k \in [0, 2\pi)$, for $V_1=V_2=0.4$, which does not encircle the EPs.  (b) The complex energy contours for (a), which form two unlinks.  (c)--(d) Similar to (a)--(b), but for $V_1=V_2=1.6$.  In this case, the parametric trajctory encircles both EPs, and the energies form a Hopf link.}
  \label{fig:figureS2}
\end{figure}

\begin{figure}
  \centering
  \includegraphics[width=\linewidth]{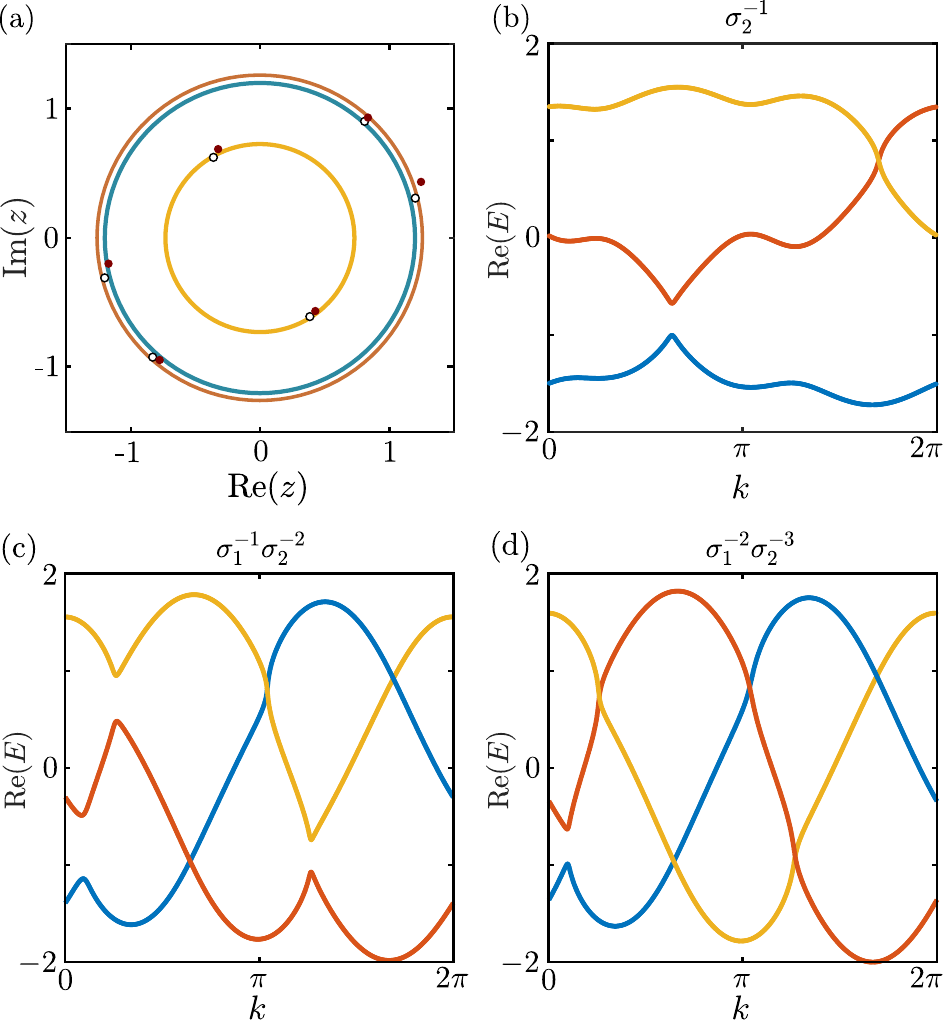}
  \caption{(a) EP positions for trimers with $V_1 = V_2$, in the parametric space of $z = V_1e^{ik}$.  The filled (hollow) dots indicate the EPs for the disordered (ordered) trimer.  The disorder consists of an additional mass term $Dp$ on each site, where $D = 1/10$ is the disorder strength and $p$ is drawn from the normal distribution.  The yellow, blue and orange parametric trajectories correspond to $V_1 = 0.72$, $V_1 = 1.2$ and $V_1 = 1.26$, respectively encircling one, three and five EPs. (b)--(d) Band diagrams for the yellow, blue and orange parametric trajectories in (a), respectively.  Each diagram is labeled by its braid word.  All other parameters are the same as in Fig.~\ref{fig:figure3}(b).}
  \label{fig:disorder}
\end{figure}

Such similarities can also be observed for the unlink and trefoil knot phases.  Larger values of $V_2$ induce interactions between the edge state sub-bands and the other eigenstates of the lattice, producing complicated braidings that lack any obvious relationship to the Hermitian AAH model.

\section{Braiding in the dimer model}
\label{secH}

We use an analysis similar to the one in Sec.~\ref{secE} to see how the unlink-to-Hopf link transition occurs.  For $M = 2$, $\alpha = 1/2$, Eqs.~\eqref{COM_AAH_Hamiltonian_1}--\eqref{Vm} produce the $2\times 2$ Hamiltonian
\begin{equation} \label{Dimer_Bulk_Hamiltonian}
  \mathcal{H}(z) =
  \begin{bmatrix}
    -z & t \\
    t & z \\
  \end{bmatrix},
\end{equation}
where $z = V_1 \mathrm{cos}(k) + iV_2 \mathrm{sin}(k)$. The eigenenergies are
\begin{equation} \label{Dimer_Bulk_Eigenvalues}
  E = \pm \sqrt{z^2+t^2}.
\end{equation}
Using $z$ as a complex (2D) parameter space, the $\pm$ solutions form two Riemann surfaces, with order-2 EPs (branch points) at $z = \pm it$.

Over one cycle of $k \in [0,2\pi)$, the $z$ parameter moves along an ellipse centered at $z = 0$ with major axes $2V_1$ and $2V_2$.  For $V_2 < t$, the parametric trajectory encircles neither EP, as shown in Fig.~\ref{fig:figureS2}(a) for $V_1 = V_2 = 0.4$.  As shown in Fig.~\ref{fig:figureS2}(b), the two energies form unlinks, and the global Berry phase is $\phi_G=0$.  For $V_2 > t$, the parametric trajectory encircles both EPs, as shown in Fig.~\ref{fig:figureS2}(c) for the case of $V_1 = V_2 = 1.6$.  As shown in Fig.~\ref{fig:figureS2}(d), Here, the energies form a Hopf link (braid word $\sigma_1^{-2}$, $\phi_G=2\pi$).

\section{Disorder-induced braid transition}
\label{appendix:EPdisorder}

As we have argued, one of the virtues of the NH AAH model, \eqref{COM_AAH_Hamiltonian_1}--\eqref{Vm}, is that many types of energy braids can be generated by varying a handful of model parameters.  As braids are discrete phenomena, they are unaffected by weak disorder in the lattice.  However, sufficiently strong disorder can induce braid transitions.

As an illustration, we return to the trimer ($M = 3$) with $V_1 = V_2$, which was previously discussed in Sec.~\ref{secE}.  We add disorder to each real on-site term by replacing $V_1\cos(2\pi\alpha m+k)$ with $V_1\cos(2\pi\alpha m+k)+ Dp$, where $D$ is a disorder strength parameter and $p$ is drawn from the normal distribution.  In the $z$-parameter space (see Sec.~\ref{secE}), the positions of the EPs are shifted by the disorder, as shown in Fig.~\ref{fig:disorder}(a).  If the shifts are small, the number of EPs encircled by the parametric trajectory (a curve of constant $|z|$) is unchanged, and the energy braids remain the same.

The EPs of the ordered trimer come in three pairs of equal $|z|$, but this symmetry is broken by the disorder.  For example, the yellow, blue and orange parametric trajectories in Fig.~\ref{fig:disorder}(a) encircles one, three and five EPs respectively.  The resulting band diagrams are shown in Fig.~\ref{fig:disorder}(b)--(d).  For instance, the complex energies in Fig.~\ref{fig:disorder}(d) form the braid word $\sigma_1^{-2}\sigma_2^{-3}$, which does not occur in the ordered version of this trimer.

\end{document}